# Towards a reliable approach on scaling in data acquisition

Nicolae PARASCHIV
*Control Engineering, Computers & Electronics Dept.*
*Petroleum-Gas University of Ploiesti*
Ploiesti, Romania
nparaschiv@upg-ploiesti.ro
https://orcid.org/0000-0002-8923-0966

Emil PRICOP
*Control Engineering, Computers & Electronics Dept.*
*Petroleum-Gas University of Ploiesti*
Ploiesti, Romania
emil.pricop@upg-ploiesti.ro
https://orcid.org/0000-0002-4021-6549

Jaouhar FATTAHI
*Computer Science & Software Engineering Dept.*
*Laval University*
Quebec City, Canada
jaouhar.fattahi.1@ulaval.ca
https://orcid.org/0000-0002-3905-9099

Florin ZAMFIR
*Control Engineering, Computers & Electronics Dept.*
*Petroleum-Gas University of Ploiesti*
Ploiesti, Romania
florin.zamfir@upg-ploiesti.ro
https://orcid.org/0000-0003-2776-9234

*Abstract*— Data acquisition is an important process in the functioning of any control system. Usually, the acquired signal is analogic, representing a continuous physical measure, and it should be processed in a digital system based on an analog to digital converter (ADC) and a microcontroller. The ADC provides the converted value in ADC units, but the system and its operator need the value expressed in physical units. In this paper we propose a novel design solution for the scaling module, which is a key component of a digital measurement system. The scaling module refers to fitting the sensor result of a variable number of bits depending on the ADC resolution into physical units. A general method for scaling is proposed and a SageMath script is presented for obtaining easily the scaling function. In the last part of the paper, the proposed method is validated in a case study, by calculus, and it is implemented on a low-cost development system in order to create a wireless sensor node.

## I. INTRODUCTION

Data acquisition is a critical process in the design and functioning of all the control systems. The reliability and correct functioning of the systems is based on the quality and precision of the measured values.

A large number of process variables are represented by continuous signals. The actual control systems are incorporating microcontrollers and embedded devices which are obviously digital. It is required to acquire the analogic value and to convert it to a digital signal, by using a standardized Analog to Digital Converter (ADC). The functioning of any ADC is well known, and it was described in [1]. The ADC output is a numeric value, representing the quantum number corresponding to the measured value, expressed in ADC units. The maximum quantum number is linked to the ADC resolution.

The ADC output, represented by quantum number, is difficult to be interpreted by a control system operator, so the conversion between ADC units and physical units is required. This conversion, that gives a physical significance to the ADC output, is realized by the Scaling Module (SM)

In this paper, a design method for the scaling module is proposed and validated. In the second section of the paper, we introduce the novel theoretical approach of the scaling module design. The third section of the paper is focused on presenting a method for the scaling function customization based on the static characteristics of the digital measurement system and its components. By customizing the function for a given system the scaling module design is validated. The fourth section is focused on presenting an implementation of a wireless sensor node using NodeMCU, a low-cost microcontroller development system. The wireless sensor node functioning is based on the software implementation of the proposed scaling module design. The authors present their conclusions and future research directions in the last section.

## II. SCALING MODULE DESIGN

The scaling module is a key component of a digital measurement system (DMS). The structure or a generic DMS is presented in Fig. 1.

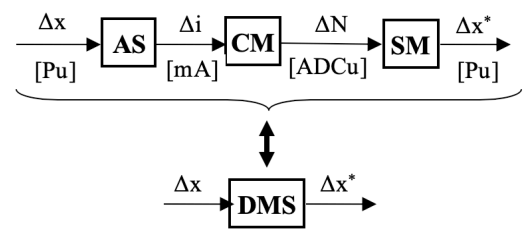

Fig. 1. Digital Measurement System structure

The DMS structure presented in Fig. 1 includes the following blocks: AS – Analogic Sensor, CM – Conversion Module and SM – Scaling Module. The evidenced signals are x – measured (acquired) parameter in Pu (Physical units), i – sensor generated current expressed in mA, N – numeric result generated by CM (ADC) expressed in ADC units – ADCu and x* - acquired parameter computed value, expressed in Physical units (Pu).

The variations of variables from reference values can be observed in Fig. 1. Given the variable $z \in [z_{\min}, z_{\max}]$, there can be defined:

- actual variation -  $\Delta z = z_{max} - z$   (1)

- maximum variation $\Delta z_{\max} = z_{\max} - z_{\min}$.   (2)

The main challenge of the scaling module (SM) design process is to find a general relationship to correlate the variation $\Delta x^*$ with $\Delta N$ variation.  The proposed scaling method is used for converting the measured values from ADC units to Physical units, respectively for implementing the F2 frontier as presented in [1].

To design the scaling module (SM) the functions associated with the static characteristics of DMS, AS and CM should be used. In the following paragraphs we considered generic static characteristics.

A generic static characteristic for DMS is represented by the function $g$, as presented in Fig. 2.

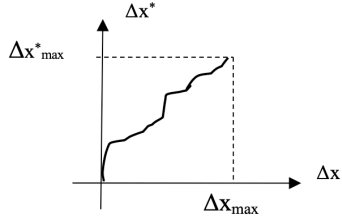

Fig. 2. The graph of $g$ function associated with the DMS.

The function:
$$\Delta x^* = g(\Delta x), \qquad (3)$$
has the following properties:

- $g$ is a continuous function;
- $g(0) = 0$;
- $g(\Delta x_{max}) = \Delta x^*_{max}$.

A generic characteristic for AS is corresponding with the function f, whose graphical representation is shown in Fig. 3.

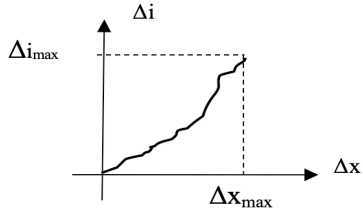

Fig. 3. The graph of $f$ function associated with the AS.

The function:
$$\Delta i = f(\Delta x), \qquad (4)$$
has the following properties:

- $f$ is a continuous function;
- $f(0) = 0$;
- $f(\Delta x_{max}) = \Delta i_{max}$.

Function f is also an invertible function, as presented in the relationship (5).
$$\Delta x = f^{-1}(\Delta i), \qquad (5)$$

The graph of the function $f^{-1}$ is presented in Fig. 4.

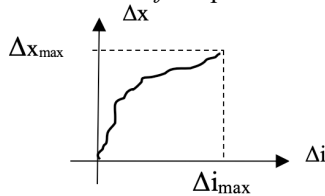

Fig. 4. The graph of $f^{-1}$ function associated with the AS.

The properties of $f^{-1}$ are listed below:
- $f^{-1}$ is a continuous function;
- $f^{-1}(0) = 0$;
- $f^{-1}(\Delta i_{max}) = \Delta x_{max}$.

A generic static characteristic of the conversion module (CM) is represented by the function h as shown in Fig. 5.

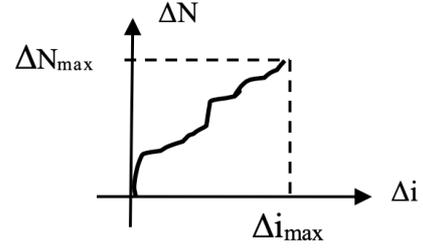

Fig. 5. The graph of $h$ function associated with MC.

The function:
$$\Delta N = h(\Delta i), \qquad (6)$$
has the following properties:

- $h$ is a continuous function;
- $h(0) = 0$;
- $h(\Delta i_{max}) = \Delta N_{max}$.

Also, $h$ is an invertible function as shown in the relationship (7).
$$\Delta i = h^{-1}(\Delta N), \qquad (7)$$

The graphical representation of function $h^{-1}$ is presented in Fig. 6.

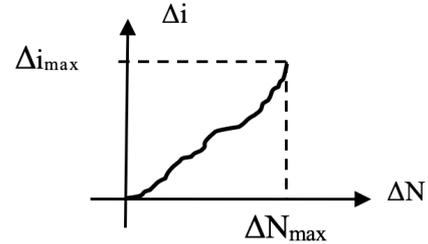

Fig. 6. The graph of $h^{-1}$ function associated with CM.

The function $h^{-1}$ has the following properties:
- $h^{-1}$ is a continuous function;
- $h^{-1}(0) = 0$;
- $h^{-1}(\Delta i_{max}) = \Delta N_{max}$.

By replacing the relationship (5) in (3), the following relationship (8) is obtained.
$$\Delta x^* = g[f^{-1}(\Delta i)], \qquad (8)$$

By taking into consideration the relationship (7) we can obtain the following relationship
$$\Delta x^* = g\{f^{-1}[h^{-1}(\Delta N)]\}, \qquad (9)$$

which can easily be replaced with:
$$\Delta x^* = q(\Delta N). \qquad (10)$$

Examining the relationships (9) and (10) in the context of Fig. 1, it can be observed that they are associated with the scaling module (SM). Moreover, starting from relation (9), the specific form of the scaling relation for different given functions $g$, $f$ and $h$ can be deduced, as it is presented in the following paper section.

### III. SCALING FUNCTION CUSTOMIZATION USING SAGEMATH. CASE STUDY

In this section of the paper, we present an example of obtaining the custom scaling function for a given measurement system. To deduce the actual form of the scaling relation, various forms of the static characteristics for the DMS, AS and CM should be used.

The provided case study is realized for a flow measurement system. The ADC used in the DMS has 10 bits resolution. The technical specifications of the DMS components are presented in Table I.

TABLE I. DMS COMPONENTS SPECIFICATIONS

| Component | Measure | Min value | Max value |
|---|---|---|---|
| Flow transducer domain | Q [m3/h] | 0 | 30 |
| Flow transducer output domain | I [mA] | 4 | 20 |
| ADC output | N [ADCu] | 0 | 1023 |

The studied DMS and its components have the following static characteristics:

- AS – non-linear, with the following static characteristic:

$$f(\Delta Q) = \frac{\Delta i_{max}}{(\Delta Q_{max})^2}\Delta Q^2 \quad (11)$$

- CM – linear, with the characteristic given by the relation:

$$h(\Delta i) = \frac{\Delta N_{max}}{\Delta i_{max}}\Delta i \quad (12)$$

- DMS – linear, with the following characteristic:

$$g(\Delta Q) = \frac{\Delta Q^*_{max}}{\Delta Q_{max}}\Delta Q \quad (13)$$

The value of the $\Delta i_{max}$, $\Delta N_{max}$, $\Delta Q_{max}$, $\Delta Q^*_{max}$, $\Delta Q$, $\Delta N$ are easily calculated by taking into consideration the graphical representations in section II of this paper and the technical specifications presented in Table I.

The calculated values are:

$$\Delta i_{max} = i_{max} - i_{min} = 20 - 4 = 16\ [mA] \quad (14)$$

$$\Delta Q_{max} = Q_{max} - Q_{min} = 30 - 0 = 30\ [m^3/h] \quad (15)$$

$$\Delta Q = Q_{max} - Q = 30 - Q\ [m^3/h] \quad (16)$$

$$N_{max} = 2^{10} - 1 = 1023\ [ADCu] \quad (17)$$

$$N_{min} = N_{max}\frac{i_{min}}{i_{max}} = 1023 * \frac{4}{20} = 204{,}6\ [ADCu] \quad (18)$$

$$\Delta N_{max} = N_{max} - N_{min} =$$
$$= 1023 - 204{,}6 = 818{,}4\ [ADCu] \quad (19)$$

$$\Delta N = N_{max} - N = 1023 - N\ [ADCu] \quad (20)$$

It should be noted that $\Delta Q^*_{max}$ and $\Delta Q_{max}$ have similar expressions.

In the following paragraphs SageMath will be used for finding the function $q$, as described in relations (10) and (9), based on the relations (11)-(20). We propose this approach in order to automate the finding process of the analytical expression of the scaling function.

SageMath [2] is an open-source, powerful and very complex mathematics software system, that integrates the flexibility of Python programming language with pre-built open-source packages as *NumPy* [3] for numerical computations, *matplotlib* [4] for rapid graphical representations, SciPy [5] for scientific computing and Maxima [6] for analytical calculations.

The SageMath software package is cross-platform, since it is based on the Python interpreter and Python specific libraries. Despite the fact it is very complex, SageMath is not very difficult to use. It has two running modes: the first is as a CLI (Command Line Interface), where an interpreter is executed in the shell of the operating system, while the second is a graphical and user-friendly mode based on Jupyter Notebooks.

The main issue to solve using SageMath is the analytical determination of the inverse function of the given functions $f$ and $g$. SageMath does not include an inverse method to call, but there can be defined an equation to be solved based on the inverse function definition, as it can be seen on lines 25 and 30 in Fig. 7.

```
1   #Compute scaling function q
2   #Variables
3   var('dNmax, dN, dimax, dQmax, dQ, dQmaxstar, dQstar')
4   var('Nmin, Nmax, imin, imax, n,  di')
5   #Sensor output domain
6   imin=4
7   imax=20
8   dimax=imax-imin
9   #Sensor measuring domain
10  Qmin=0
11  Qmax=30
12  dQmax=Qmax-Qmin
13  dQmaxstar=Qmax-Qmin
14  #ADC parameters
15  #n – resolution
16  n=10
17  Nmax=2^n-1
18  Nmin=Nmax*imin/imax
19  dNmax=Nmax-Nmin
20  #computing of the characteristic functions
21  #computing h function (CM characteristic)
22  h(di)=dNmax*di/dimax
23  #computing of inverse h function
24  var('y1');
25  hinv(dN)=solve(dN==h(y1),y1)[0].rhs()
26  #computing f function (AS characteristic)
27  f(dQ)=dimax*dQ*dQ/(dQmax*dQmax)
28  #computing of inverse f function
29  var('y2')
30  finv(dQ)=solve(dQ==f(y2),y2)[1].rhs()
31  #computing of g function (DMS characteristic)ș
32  g(dQ)=dQmaxstar/dQmax*dQ
33  #computing the q function (SM characteristic)
34  q(dN)=g(finv(hinv(dN)))
35  #printing the result on screen
36  print q
```

Fig. 7. SageMath script code for computing the Scaling Function $q$

In Fig. 7 is presented the complete SageMath script for computing the q function. The variables values are assigned in the first part of the script (lines 6 – 19). The functions $h$, $h^{-1}$, $f$, $f^{-1}$, and $g$ are calculated symbolically in the second part of the program, as it is observed on lines 22-32. The calculated functions are not displayed to the user. The $q$ function (Scaling

Module characteristic) is calculated based on therelation (9) using the instruction on line 34, and the result is displayed on the screen.

The program is stored as *compute_q.sage* script that can be run in a Terminal or Command prompt. In Fig. 8, there are presented the results of running the developed scrip. In the second row the scaling function expression can be observed.

```
Emils-MacBook-Pro:Desktop emilpricop$ sage compute_q.sage
dN |--> 15*sqrt(5/1023)*sqrt(dN)
```

Fig. 8. SageMath script results – the expression of the scaling function for the considered DMS, as described in Table I)

By running the SageMath script, the final expression for the scaling function is:

$$\Delta Q^* = 15\sqrt{\frac{5}{1023}}\sqrt{\Delta N} \quad (21)$$

By taking into account the variation values and computing the algebraic operations from (21) the following relation is obtained:

$$Q^* - Q^*_{min} = 1,04866903495\sqrt{N - N_{min}} \quad (22)$$

The final expression of $Q^*$ (23) is obtained by replacing the numerical values in (22).

$$Q^* = 1,04866903495\sqrt{N - 204,6} \quad (23)$$

Equation (23) can be used for implementing the scaling module and its characteristic scaling function on an embedded device or microcontroller, as it is presented in the following section.

## IV. EMBEDDED SYSTEM-BASED IMPLEMENTATION OF THE SCALING FUNCTION

The results obtained in the second and third sections of this paper can be used for implementing a software scaling module on an embedded system. This approach can be used for implementing a rapid operating wireless sensor node, which reads an analog signal, converts it to a digital value and transmits the value by a radio link to a data concentrator.

The implementation of a simple wireless sensor node using an embedded system is presented in this section of the paper. The platform used for the experimental implementation is NodeMCU. This low-cost development platform was becoming very popular in the last two years due to its very low price and open-hardware and open-firmware design.

The core of NodeMCU is ESP8266, a Wi-Fi system-on-chip (SoC) produced by Chinese company EspressIF Systems. The SoC contains an 80 MHz 32-bit processor, 512 KB RAM memory, 512 KB Flash memory, IEEE 802.11 b/g/n compatible Wi-Fi module and one analog to digital converter (ADC) with 10-bit resolution [7].

The most important feature of NodeMCU, apart from its price, which is around 7 USD for mass markets, is the very low current consumption 10 μA – 170 mA. Taking into consideration this technical characteristic, we can consider NodeMCU to be a good candidate for developing a sensor node that can be powered by a battery or by a solar panel. The node will send the measured value only at preconfigured intervals, for example every 60 seconds.

The flow transducer having the parameters presented in the previous paper section is connected to the analog input of the NodeMCU board as presented in Fig. 9.

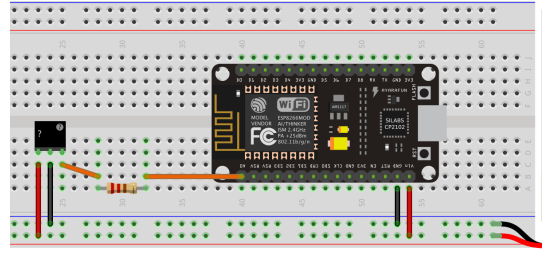

Fig. 9. Connection diagram of the sensor to the A0 pin of NodeMCU

Since the transducer output is in the 4 … 20 mA domain, a 165Ω resistor should be used to provide the voltage in the domain accepted by the NodeMCU analog input, respectively 0…3,3V.

The sensor node should provide the user with the flow value measured in physical units. This value can easily be obtained by using relationship (23) and computing it for a given N – number of quanta resulted from the ADC.

In order to increase the power saving capability and to allow rapid calculation of the measured value expressed in physical units we propose an alternative method for this calculation, by using the final form of the relation (23), initially calculated with the SageMath Script. The logical flow of the proposed program is presented in Fig. 10.

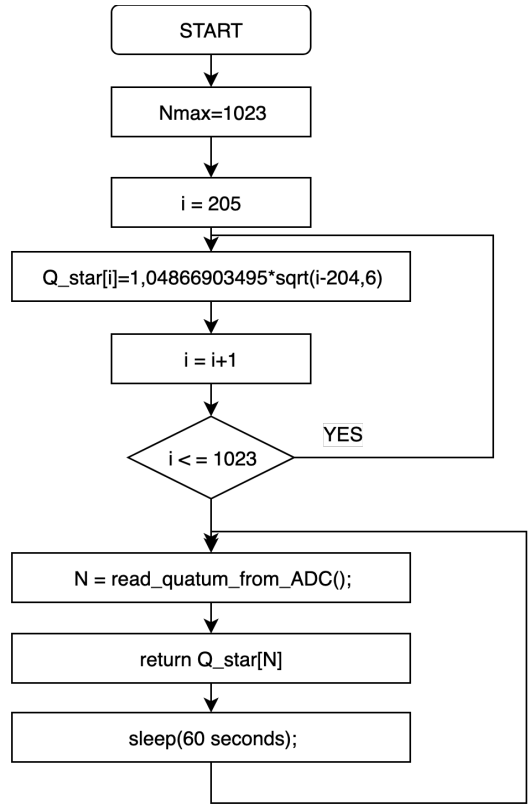

Fig. 10. Scaling function calculation program logical flow diagram for implementation on NodeMCU development system (10-bit ADC)

Basically, the program has two fundamental parts: an initialization zone and the value reading and transmission routine.

When the NodeMCU development platform starts-up (boot-up) the program initializes a unidimensional array (Q_star) with the index from 205 to 1023 with the calculated value of relationship (23). The unidimensional array will be in this case a real conversion table: the index of the array represents the number of quanta expressed in ADC units, and the value of the array element represents the measured value in physical units. For example, Q_star[1023]=30 and Q_star[418]=15.

After initializing the array, the NodeMCU program enters an infinite loop where the microcontroller reads the sensor provided analog value at every minute. The conversion is pursued by the ADC and the program receives the N number of quanta. By using the unidimensional array Q_star, the measured value expressed in physical units can be easily found and transmitted to the remote data logger. The NodeMCU board enters the low-power consumption (sleep) mode for 60 seconds.

The C code for implementing the program on NodeMCU board is presented in Fig. 11.

```
double Q_star[1024]; // unidimensional array - conversion table
const int analogInPin = A0; // Analog input pin of the NodeMCU
double outputValue = 0;
int N = 205; // initial temporary variable
void setup() {
  //initialization zone
  int i;
  for (i=205; i<=1023; i++)
  {
    //array initialization with the conversion table
    Q_star[i]= 1.04866903495*sqrt(i-204.6);
  }
}
void send_data(float out)
{
  //code for wireless data sending or printing on serial console
}
void loop() {
  //infinite loop
  //reading measured value from ADC in ADC units
  N=analogRead(analogInPin);
  //conversion from ADC units to physical units
  outputValue=Q_star[N];
  //routine to send data to the datalogger or to serial console
  send_data(outputValue);
  //put the NodeMCU board in sleep mode for 60 seconds
  ESP.deepSleep(60e6);
}
```

Fig. 11. The source code for NodeMCU programming

As it can be observed in Fig. 10 and Fig. 11 the program calculates the $Q^*$ values for $N \in [205, 1023]$. The ADC output is restricted to this interval in the case of the studied system. N=205 ADCu is the rounded value corresponding to the minimum measured flow and N=1023 ADCu corresponds to the maximum flow for the transducer domain.

In Table II, some representative values computed using the identified function $q$ and the proposed program are presented.

The NodeMCU is programmed using the Arduino Studio IDE. This programming software supports C programming and console debugging of the solution. Also, it provides valuable information about memory usage. According to Arduino Studio information the proposed solution uses 24% of NodeMCU 1.0 (ESP-12E based) program storage space and about 50% of the dynamic memory for storing the declared variables.

TABLE II. SAMPLE VALUES COMPUTED WITH THE PROGRAM

| N [ADCu] | $Q^*$ [$m^3$/h] |
|---|---|
| 205 | 0,66 |
| 206 | 1,24 |
| 207 | 1,62 |
| … | … |
| 1020 | 29,94 |
| 1021 | 29,96 |
| 1022 | 29,98 |
| 1023 | 30,00 |

## V. CONCLUSION

In this paper we propose a general relation that underlies the synthesis of the conversion module in ADC units in physical units within a DMS. The usage implementation of the general relation is detailed and exemplified by using a flow measurement system.

The implementation is realized in two distinct steps. The first is pursued using the SageMath open-source package and aims to find the static characteristic (function) for the scaling module. The second step consists of programming the identified function on a microcontroller.

By implementing the scaling module in software on a microcontroller with Wi-Fi communication capabilities, for example on NodeMCU board, an intelligent wireless sensor node can be created.

The obtained results can be extended, and future research direction aims to inverse conversion from physical units to DAC units, in order to be implemented for command sending and distribution devices.